\documentclass[aps,prb,superscriptaddress,reprint]{revtex4-1}
\usepackage{graphicx}
\usepackage{dcolumn}
\usepackage{bm}
\usepackage[colorlinks,linkcolor=blue,citecolor=blue,urlcolor=blue]{hyperref}
\usepackage{amsmath,amsthm,amssymb}
\usepackage{amsfonts}
\usepackage{icomma}
\usepackage{epstopdf}
\usepackage{placeins}
\usepackage{color}
\usepackage[euler]{textgreek}
\usepackage[separate-uncertainty = true, exponent-product = \times]{siunitx}
\graphicspath{{./Plots/}}

\makeatletter
\let\cat@comma@active\@empty
\makeatother

\begin{document}

\title{Magnetic Tunnel Junction Performance Under Mechanical Strain}

\author{Niklas Roschewsky}
\email{roschewsky@berkeley.edu}
\affiliation{Department of Physics, University of California, Berkeley, California 94720, USA}

\author{Sebastian Schafer}
\affiliation{Samsung Semiconductor Inc., NMT, 3655 N 1st St, San Jose, CA-95134}

\author{Frances Hellman}
\affiliation{Department of Physics, University of California, Berkeley, California 94720, USA}

\author{Vladimir Nikitin}
\email{v.nikitin@samsung.com}
\affiliation{Samsung Semiconductor Inc., NMT, 3655 N 1st St, San Jose, CA-95134}

\date{\today}

\begin{abstract}
In this work we investigate the effect of the mechanical stress on the performance of magnetic tunnel junctions (MTJ) with perpendicular magnetic anisotropy. We developed a 4-point bending setup, that allows us to apply a constant stress over a large substrate area with access to electrical measurements and external magnetic field. This setup enables us to measure key device performance parameters, such as tunnel magnetoresistance (TMR), switching current ($I_c^{50\%}$) and thermal stability ($\Delta$), as a function of applied stress. We find that variations in these parameters are negligible: less than $\SI{2}{\percent}$ over the entire measured range between the zero stress condition and the maximum stress at the point of wafer breakage.
\end{abstract}

\maketitle

Recently, several companies have announced successful integration of embedded Magnetic Random Access Memory (MRAM) with existing CMOS logic~\cite{Kang2014,Shum2017,Song2016}. Spin transfer torque (STT) MRAM is a non-volatile memory technology that offers high speeds, low energy consumption and high endurance~\cite{Apalkov2016,Kawahara2012}. The fundamental building block of STT-MRAM is a magnetic tunnel junction (MTJ), which consists of two ferromagnetic layers separated by a tunneling barrier. Readout of the MRAM bit is enabled by the tunnel magnetoresistance (TMR) of the MTJ~\cite{Julliere1975,Moodera1995}, while write operations are based on STT switching~\cite{Berger1996,Slonczewski1996}.\\
An important design parameter for MRAM is the strain applied to the ferromagnetic layers of the MTJ. Strain can impact the magnetic and electronic properties of a magnet as well as the quantum transport across the tunneling barrier. The TMR of a MTJ with in-plane anisotropy changes significantly under application of stress~\cite{Loong2016}. In fact, Loong et al. have seen an enhancement of the TMR by \SI{68}{\percent}~\cite{Loong2015} under the application of inhomogeneous strain. Furthermore, strain and pressure sensors based on the magneto elastic coupling of CoFe have been demonstrated~\cite{Lohndorf2002,Tavassolizadeh2015,Tavassolizadeh2016,Meyners2009}.\\
Previous work on the strain dependence of MTJs has focused on devices with in-plane magnetic anisotropy. State of the art memory elements, however, utilize MTJs having thinner free layers with out of plane anisotropy due to better scalability and faster switching times~\cite{Sun2000,Worledge2011,Nowak2011,Sato2013}. \\
In this work we characterize MTJs with out of plane magnetic anisotropy under systematic application of strain. In addition to the TMR, we also study the strain dependence of other important performance parameters, such as the critical write current $I_c^{50\%}$ and the thermal stability factor $\Delta$. To apply the strain in a systematic way, we have designed an integrated 4-point bending setup~\cite{Baril1999} with a magnetic probe station. This 4-point bending setup allows us to apply constant strain over large substrate areas while magnetotransport measurements are carried out.\\
We present the surprising result that transport in our MTJ devices with perpendicular magnetic anisotropy is very robust to mechanical stress. Our findings show that while the TMR and the thermal stability factor $\Delta$ are independent of external strain (within the accuracy of the measurement), we observe a small decrease of the coercive field $\mu_0 H_c$ and the switching current $I_c^{50\%}$ with increasing strain.\\
\begin{figure}
\begin{center}
\includegraphics[width=\columnwidth]{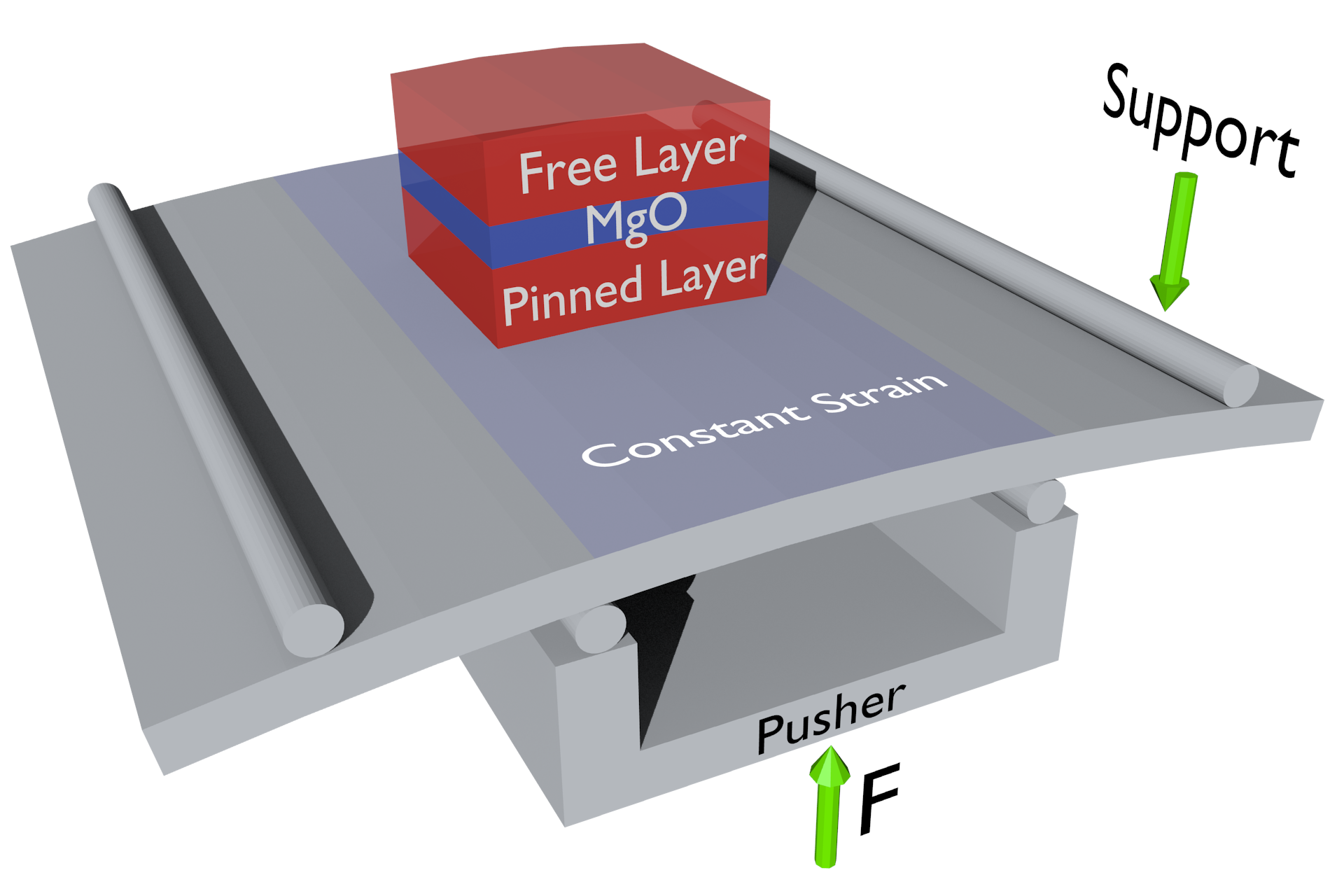}
\caption{\label{fig:Fig1} Schematics of the 4-point bending setup. The wafer is pushed up by a force F in the center while two supporting pins hold the position fixed at the outside edges. The mechanical stress in the central region (light gray) is constant. The magnetic tunneljunctions under test have perpendicular magnetic anisotropy.}
\end{center}
\end{figure}
Figure~\ref{fig:Fig1} shows a sketch of our 4-point bending setup. Two ceramic rods support the wafer from the top while a force is applied from the bottom to bend the wafer. The strain $\epsilon$ in the center region between the two supporting rods is constant. This strain is related to the bending curvature $\kappa$ as $\epsilon=\kappa\cdot y$, where $y$ is the distance to the neutral plane, i.e the wafer thickness divided by 2.\\
The MTJs under investigation in this study are deposited on top of a \SI{775}{\micro\meter} thick silicon wafer with \SI{100}{\nano\meter} thermal oxide. The MTJ stack consists of a synthetic antiferromagnet, i.e. layered ferromagnets with antiferromagnetic coupling, serving as a reference (pinned) layer, and a CoFeB-based free layer separated by a MgO tunnel barrier. Both layers have perpendicular magnetic anisotropy. Tunnel junction devices are patterned, using conventional ion milling technique, into circles with diameter $d$ from \SIrange{40}{80}{\nano\meter}. All measurements of the electrical resistance $R$ are performed in a two terminal geometry with the contact resistance taken into account.\\
To protect our samples from damage due to uncontrolled shattering we first examine the maximum strain that can be applied to the wafer before a catastrophic breakage event occurs. We find that the average breaking point of the silicon wafers used for this study is at $\epsilon \approx \SI{0.1}{\percent}$. Assuming the value of $Y=\SI{150}{\giga\pascal}$ for the Young's modulus of silicon, this corresponds to a stress level of $\sigma=\SI{150}{\mega\pascal}$. To prevent wafer breakage, the maximum strain applied in this study is thus limited to $\SI{0.06}{\percent}$.\\
\begin{figure}
\begin{center}
\begingroup
  \makeatletter
  \providecommand\color[2][]{%
    \GenericError{(gnuplot) \space\space\space\@spaces}{%
      Package color not loaded in conjunction with
      terminal option `colourtext'%
    }{See the gnuplot documentation for explanation.%
    }{Either use 'blacktext' in gnuplot or load the package
      color.sty in LaTeX.}%
    \renewcommand\color[2][]{}%
  }%
  \providecommand\includegraphics[2][]{%
    \GenericError{(gnuplot) \space\space\space\@spaces}{%
      Package graphicx or graphics not loaded%
    }{See the gnuplot documentation for explanation.%
    }{The gnuplot epslatex terminal needs graphicx.sty or graphics.sty.}%
    \renewcommand\includegraphics[2][]{}%
  }%
  \providecommand\rotatebox[2]{#2}%
  \@ifundefined{ifGPcolor}{%
    \newif\ifGPcolor
    \GPcolortrue
  }{}%
  \@ifundefined{ifGPblacktext}{%
    \newif\ifGPblacktext
    \GPblacktextfalse
  }{}%
  \let\gplgaddtomacro\g@addto@macro
  \gdef\gplbacktext{}%
  \gdef\gplfronttext{}%
  \makeatother
  \ifGPblacktext
    \def\colorrgb#1{}%
    \def\colorgray#1{}%
  \else
    \ifGPcolor
      \def\colorrgb#1{\color[rgb]{#1}}%
      \def\colorgray#1{\color[gray]{#1}}%
      \expandafter\def\csname LTw\endcsname{\color{white}}%
      \expandafter\def\csname LTb\endcsname{\color{black}}%
      \expandafter\def\csname LTa\endcsname{\color{black}}%
      \expandafter\def\csname LT0\endcsname{\color[rgb]{1,0,0}}%
      \expandafter\def\csname LT1\endcsname{\color[rgb]{0,1,0}}%
      \expandafter\def\csname LT2\endcsname{\color[rgb]{0,0,1}}%
      \expandafter\def\csname LT3\endcsname{\color[rgb]{1,0,1}}%
      \expandafter\def\csname LT4\endcsname{\color[rgb]{0,1,1}}%
      \expandafter\def\csname LT5\endcsname{\color[rgb]{1,1,0}}%
      \expandafter\def\csname LT6\endcsname{\color[rgb]{0,0,0}}%
      \expandafter\def\csname LT7\endcsname{\color[rgb]{1,0.3,0}}%
      \expandafter\def\csname LT8\endcsname{\color[rgb]{0.5,0.5,0.5}}%
    \else
      \def\colorrgb#1{\color{black}}%
      \def\colorgray#1{\color[gray]{#1}}%
      \expandafter\def\csname LTw\endcsname{\color{white}}%
      \expandafter\def\csname LTb\endcsname{\color{black}}%
      \expandafter\def\csname LTa\endcsname{\color{black}}%
      \expandafter\def\csname LT0\endcsname{\color{black}}%
      \expandafter\def\csname LT1\endcsname{\color{black}}%
      \expandafter\def\csname LT2\endcsname{\color{black}}%
      \expandafter\def\csname LT3\endcsname{\color{black}}%
      \expandafter\def\csname LT4\endcsname{\color{black}}%
      \expandafter\def\csname LT5\endcsname{\color{black}}%
      \expandafter\def\csname LT6\endcsname{\color{black}}%
      \expandafter\def\csname LT7\endcsname{\color{black}}%
      \expandafter\def\csname LT8\endcsname{\color{black}}%
    \fi
  \fi
    \setlength{\unitlength}{0.0500bp}%
    \ifx\gptboxheight\undefined%
      \newlength{\gptboxheight}%
      \newlength{\gptboxwidth}%
      \newsavebox{\gptboxtext}%
    \fi%
    \setlength{\fboxrule}{0.5pt}%
    \setlength{\fboxsep}{1pt}%
\begin{picture}(4874.00,5102.00)%
    \gplgaddtomacro\gplbacktext{%
      \csname LTb\endcsname%
      \put(792,3255){\makebox(0,0)[r]{\strut{}}}%
      \put(792,3651){\makebox(0,0)[r]{\strut{}}}%
      \put(792,4046){\makebox(0,0)[r]{\strut{}}}%
      \put(792,4442){\makebox(0,0)[r]{\strut{}}}%
      \put(792,4837){\makebox(0,0)[r]{\strut{}}}%
      \put(924,3035){\makebox(0,0){\strut{}$-100$}}%
      \put(1713,3035){\makebox(0,0){\strut{}$-50$}}%
      \put(2503,3035){\makebox(0,0){\strut{}$0$}}%
      \put(3292,3035){\makebox(0,0){\strut{}$50$}}%
      \put(4081,3035){\makebox(0,0){\strut{}$100$}}%
      \csname LTb\endcsname%
      \put(49,4846){\makebox(0,0)[l]{\strut{}(a)}}%
    }%
    \gplgaddtomacro\gplfronttext{%
      \csname LTb\endcsname%
      \put(550,4046){\rotatebox{-270}{\makebox(0,0){\strut{}$R$}}}%
      \put(2502,2705){\makebox(0,0){\strut{}$\mu_0H$ (mT)}}%
    }%
    \gplgaddtomacro\gplbacktext{%
      \colorrgb{0.00,0.38,0.68}%
      \put(792,704){\makebox(0,0)[r]{\strut{}$0.99$}}%
      \colorrgb{0.00,0.38,0.68}%
      \put(792,1496){\makebox(0,0)[r]{\strut{}$1$}}%
      \colorrgb{0.00,0.38,0.68}%
      \put(792,2287){\makebox(0,0)[r]{\strut{}$1.01$}}%
      \csname LTb\endcsname%
      \put(924,484){\makebox(0,0){\strut{}$0$}}%
      \put(1826,484){\makebox(0,0){\strut{}$0.02$}}%
      \put(2728,484){\makebox(0,0){\strut{}$0.04$}}%
      \put(3630,484){\makebox(0,0){\strut{}$0.06$}}%
      \colorrgb{0.87,0.09,0.12}%
      \put(4213,704){\makebox(0,0)[l]{\strut{}$56$}}%
      \colorrgb{0.87,0.09,0.12}%
      \put(4213,1337){\makebox(0,0)[l]{\strut{}$58$}}%
      \colorrgb{0.87,0.09,0.12}%
      \put(4213,1970){\makebox(0,0)[l]{\strut{}$60$}}%
      \csname LTb\endcsname%
      \put(49,2295){\makebox(0,0)[l]{\strut{}(b)}}%
    }%
    \gplgaddtomacro\gplfronttext{%
      \colorrgb{0.00,0.38,0.68}%
      \put(220,1495){\rotatebox{-270}{\makebox(0,0){\strut{}norm. TMR}}}%
      \colorrgb{0.87,0.09,0.12}%
      \put(4652,1495){\rotatebox{-270}{\makebox(0,0){\strut{}$\mu_0 H_c$ (mT)}}}%
      \csname LTb\endcsname%
      \put(2502,154){\makebox(0,0){\strut{}$\epsilon $ ($\%$)}}%
    }%
    \gplbacktext
    \put(0,0){\includegraphics{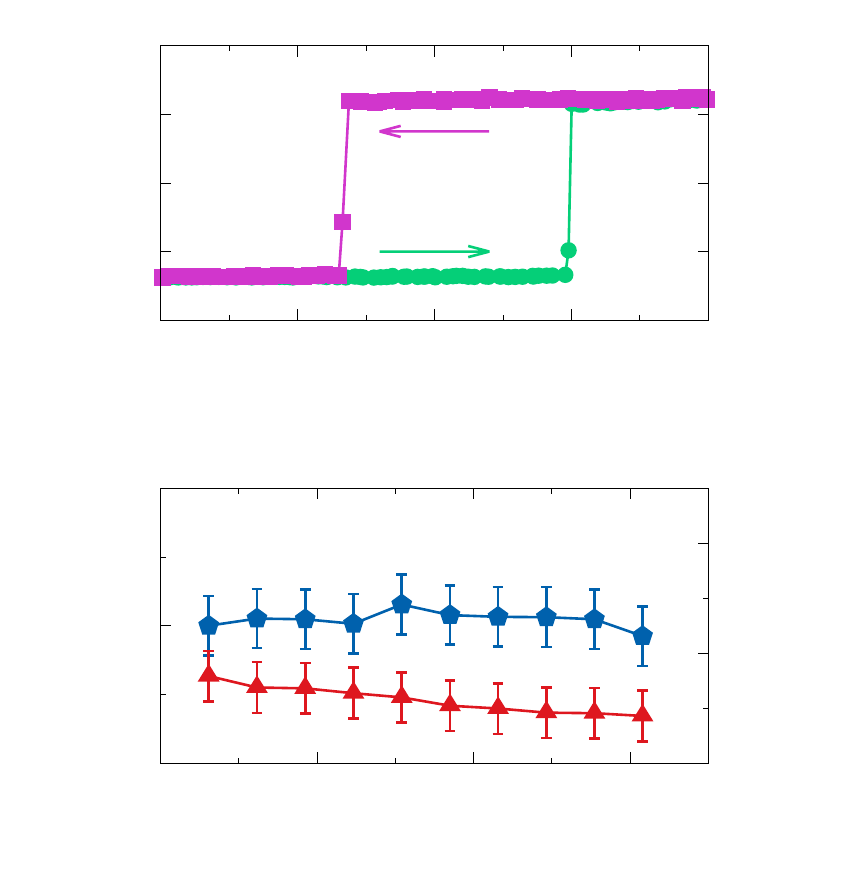}}%
    \gplfronttext
  \end{picture}%
\endgroup
\caption{\label{fig:Fig2} (a) RH minor loop for a typical magnetic tunnel junction (MTJ) under test. Arrows indicate the sweep direction. (b) Tunnel magnetoresistance TMR (left) and coercive field $\mu_0H_c$ (right) as a function of strain $\epsilon$. The TMR is constant with variations less than 1\%. The coercive field decreases by less than \SI{1}{\milli\tesla} over the full measurement range.}
\end{center}
\end{figure}
We first measure $RH$-minor loops by sweeping an external magnetic field along the easy axis of the free layer. Figure~\ref{fig:Fig2}(a) shows two switching events and clearly distinguishable high and low resistance states. The coercive field of the free layer $\mu_0H_c$ as well as the TMR ratio are extracted from this measurement. A summary of the normalized TMR ratio and $\mu_0H_c$ as a function of strain is shown on Fig.~\ref{fig:Fig2}(b). Each value reflects the mean of 46 $RH$-loops on 63 different devices. The obtained values for the TMR ratio are remarkably constant over the whole range of strain, with variations less than $\SI{1}{\percent}$. The coercive field decreases slightly with increasing strain. This decrease is attributed to the magnetoelastic coupling of the free layer.\\
\begin{figure}
\begin{center}
\begingroup
  \makeatletter
  \providecommand\color[2][]{%
    \GenericError{(gnuplot) \space\space\space\@spaces}{%
      Package color not loaded in conjunction with
      terminal option `colourtext'%
    }{See the gnuplot documentation for explanation.%
    }{Either use 'blacktext' in gnuplot or load the package
      color.sty in LaTeX.}%
    \renewcommand\color[2][]{}%
  }%
  \providecommand\includegraphics[2][]{%
    \GenericError{(gnuplot) \space\space\space\@spaces}{%
      Package graphicx or graphics not loaded%
    }{See the gnuplot documentation for explanation.%
    }{The gnuplot epslatex terminal needs graphicx.sty or graphics.sty.}%
    \renewcommand\includegraphics[2][]{}%
  }%
  \providecommand\rotatebox[2]{#2}%
  \@ifundefined{ifGPcolor}{%
    \newif\ifGPcolor
    \GPcolortrue
  }{}%
  \@ifundefined{ifGPblacktext}{%
    \newif\ifGPblacktext
    \GPblacktextfalse
  }{}%
  \let\gplgaddtomacro\g@addto@macro
  \gdef\gplbacktext{}%
  \gdef\gplfronttext{}%
  \makeatother
  \ifGPblacktext
    \def\colorrgb#1{}%
    \def\colorgray#1{}%
  \else
    \ifGPcolor
      \def\colorrgb#1{\color[rgb]{#1}}%
      \def\colorgray#1{\color[gray]{#1}}%
      \expandafter\def\csname LTw\endcsname{\color{white}}%
      \expandafter\def\csname LTb\endcsname{\color{black}}%
      \expandafter\def\csname LTa\endcsname{\color{black}}%
      \expandafter\def\csname LT0\endcsname{\color[rgb]{1,0,0}}%
      \expandafter\def\csname LT1\endcsname{\color[rgb]{0,1,0}}%
      \expandafter\def\csname LT2\endcsname{\color[rgb]{0,0,1}}%
      \expandafter\def\csname LT3\endcsname{\color[rgb]{1,0,1}}%
      \expandafter\def\csname LT4\endcsname{\color[rgb]{0,1,1}}%
      \expandafter\def\csname LT5\endcsname{\color[rgb]{1,1,0}}%
      \expandafter\def\csname LT6\endcsname{\color[rgb]{0,0,0}}%
      \expandafter\def\csname LT7\endcsname{\color[rgb]{1,0.3,0}}%
      \expandafter\def\csname LT8\endcsname{\color[rgb]{0.5,0.5,0.5}}%
    \else
      \def\colorrgb#1{\color{black}}%
      \def\colorgray#1{\color[gray]{#1}}%
      \expandafter\def\csname LTw\endcsname{\color{white}}%
      \expandafter\def\csname LTb\endcsname{\color{black}}%
      \expandafter\def\csname LTa\endcsname{\color{black}}%
      \expandafter\def\csname LT0\endcsname{\color{black}}%
      \expandafter\def\csname LT1\endcsname{\color{black}}%
      \expandafter\def\csname LT2\endcsname{\color{black}}%
      \expandafter\def\csname LT3\endcsname{\color{black}}%
      \expandafter\def\csname LT4\endcsname{\color{black}}%
      \expandafter\def\csname LT5\endcsname{\color{black}}%
      \expandafter\def\csname LT6\endcsname{\color{black}}%
      \expandafter\def\csname LT7\endcsname{\color{black}}%
      \expandafter\def\csname LT8\endcsname{\color{black}}%
    \fi
  \fi
    \setlength{\unitlength}{0.0500bp}%
    \ifx\gptboxheight\undefined%
      \newlength{\gptboxheight}%
      \newlength{\gptboxwidth}%
      \newsavebox{\gptboxtext}%
    \fi%
    \setlength{\fboxrule}{0.5pt}%
    \setlength{\fboxsep}{1pt}%
\begin{picture}(4874.00,5102.00)%
    \gplgaddtomacro\gplbacktext{%
      \csname LTb\endcsname%
      \put(792,3255){\makebox(0,0)[r]{\strut{}$-10$}}%
      \put(792,3782){\makebox(0,0)[r]{\strut{}$-5$}}%
      \put(792,4310){\makebox(0,0)[r]{\strut{}$0$}}%
      \put(792,4837){\makebox(0,0)[r]{\strut{}$5$}}%
      \put(924,3035){\makebox(0,0){\strut{}$10^{1}$}}%
      \put(1880,3035){\makebox(0,0){\strut{}$10^{2}$}}%
      \put(2837,3035){\makebox(0,0){\strut{}$10^{3}$}}%
      \put(3793,3035){\makebox(0,0){\strut{}$10^{4}$}}%
      \csname LTb\endcsname%
      \put(49,4846){\makebox(0,0)[l]{\strut{}(a)}}%
    }%
    \gplgaddtomacro\gplfronttext{%
      \csname LTb\endcsname%
      \put(181,4046){\rotatebox{-270}{\makebox(0,0){\strut{}$I_{50\%}$ (a. u.)}}}%
      \put(2502,2705){\makebox(0,0){\strut{}$t_\text{PW}$ (ns)}}%
      \colorrgb{0.02,0.81,0.47}%
      \put(3293,3870){\makebox(0,0)[l]{\strut{}p $\rightarrow$ a-p}}%
      \colorrgb{0.82,0.21,0.80}%
      \put(3293,4573){\makebox(0,0)[l]{\strut{}a-p $\rightarrow$ p}}%
      \colorrgb{0.40,0.40,0.00}%
      \put(1500,3431){\makebox(0,0)[l]{\strut{}precessional}}%
      \colorrgb{0.00,0.40,0.40}%
      \put(3293,3431){\makebox(0,0)[l]{\strut{}thermal}}%
    }%
    \gplgaddtomacro\gplbacktext{%
      \csname LTb\endcsname%
      \put(792,836){\makebox(0,0)[r]{\strut{}-5.42}}%
      \put(792,1166){\makebox(0,0)[r]{\strut{}-5.33}}%
      \put(792,1496){\makebox(0,0)[r]{\strut{}}}%
      \put(792,1825){\makebox(0,0)[r]{\strut{}2.83}}%
      \put(792,2155){\makebox(0,0)[r]{\strut{}2.92}}%
      \put(924,484){\makebox(0,0){\strut{}$0$}}%
      \put(1826,484){\makebox(0,0){\strut{}$0.02$}}%
      \put(2728,484){\makebox(0,0){\strut{}$0.04$}}%
      \put(3630,484){\makebox(0,0){\strut{}$0.06$}}%
      \colorrgb{0.87,0.09,0.12}%
      \put(1150,1265){\makebox(0,0)[l]{\strut{}parallel $\rightarrow$ anti-parallel}}%
      \colorrgb{0.00,0.38,0.68}%
      \put(1150,1627){\makebox(0,0)[l]{\strut{}anti-parallel $\rightarrow$ parallel}}%
      \csname LTb\endcsname%
      \put(49,2295){\makebox(0,0)[l]{\strut{}(b)}}%
    }%
    \gplgaddtomacro\gplfronttext{%
      \csname LTb\endcsname%
      \put(220,1495){\rotatebox{-270}{\makebox(0,0){\strut{}$I_{50\%}^{100\text{ ns}}$ (a. u.)}}}%
      \put(2502,154){\makebox(0,0){\strut{}$\epsilon $ ($\%$)}}%
    }%
    \gplbacktext
    \put(0,0){\includegraphics{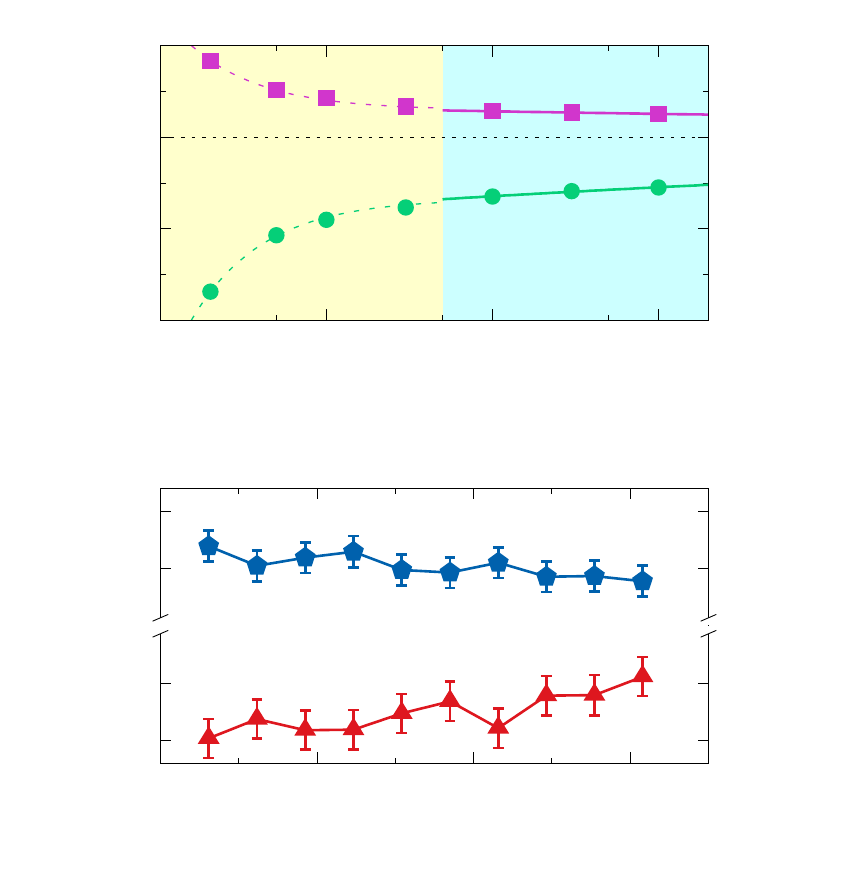}}%
    \gplfronttext
  \end{picture}%
\endgroup
\caption{\label{fig:Fig3} (a) Measurement of the critical switching current $I_{50\%}$ as a function of pulse width $t_{PW}$. $I_{50\%}$ is defined as the current where the switching probability is $\SI{50}{\percent}$. Solid lines indicate fits to a thermal activation model. At short pulses with $t_\text{PW}<\SI{500}{\nano\second}$, precessional switching is observed (dotted line). (b) shows the switching current at $\SI{100}{\nano\second}$ pulse width $I_{50\%}^{\SI{100}{\nano\second}}$ as a function of strain. For both switching directions, $I_{50\%}^{\SI{100}{\nano\second}}$ is reduced at higher strain.}
\end{center}
\end{figure}
Next, the critical current for spin-transfer torque switching is considered. We define the critical current $I_c^{50\%}$ as the current with 50\% switching probability. Figure~\ref{fig:Fig3}(a) shows $I_{50\%}$ as a function of the current pulse width $t_\text{PW}$. At long current pulses ($t_\text{PW}>\SI{500}{\nano\second}$), the switching process is thermally activated (cyan region) and $I_{50\%}$ depends logarithmically on $t_\text{PW}$~\cite{Baril1999}:
\begin{align*}
I_c = I_{c0}\left[1-\sqrt{\frac{1}{\Delta}\log\frac{t_\text{PW}}{\tau_0}}\,\right]\,.
\end{align*}
Here $\Delta$ is the thermal stability, $\tau_0$ is the intrinsic attempt time and $I_{c0}$ is the intrinsic switching current. The solid lines in Fig.~\ref{fig:Fig3}(a) are fits to the model above. For short pulses with $t_\text{PW}<\SI{500}{\nano\second}$, precessional switching is observed, where $I_c\propto 1/t_\text{PW}$~\cite{Worledge2011}. The dotted line in the yellow regime in Fig.~\ref{fig:Fig3}(a) indicates this inverse trend. \\
The critical current at $t_\text{PW}=\SI{100}{\nano\second}$, $I_{50\%}^{\SI{100}{\nano\second}}$, is shown as a function of strain in Fig.~\ref{fig:Fig3}(b). As in Fig~\ref{fig:Fig2}(b) we show the average value for 63 tested devices. For both switching directions, $I_{50\%}^{\SI{100}{\nano\second}}$ decreases with increasing strain. At $\SI{0.06}{\percent}$ strain, $I_{50\%}^{\SI{100}{\nano\second}}$ is reduced by $\approx\SI{1.5}{\percent}$. The decrease in $I_{50\%}^{\SI{100}{\nano\second}}$ is similar to the reduction in $\mu_0H_c$ with increasing strain.\\
\begin{figure}
\begin{center}
\begingroup
  \makeatletter
  \providecommand\color[2][]{%
    \GenericError{(gnuplot) \space\space\space\@spaces}{%
      Package color not loaded in conjunction with
      terminal option `colourtext'%
    }{See the gnuplot documentation for explanation.%
    }{Either use 'blacktext' in gnuplot or load the package
      color.sty in LaTeX.}%
    \renewcommand\color[2][]{}%
  }%
  \providecommand\includegraphics[2][]{%
    \GenericError{(gnuplot) \space\space\space\@spaces}{%
      Package graphicx or graphics not loaded%
    }{See the gnuplot documentation for explanation.%
    }{The gnuplot epslatex terminal needs graphicx.sty or graphics.sty.}%
    \renewcommand\includegraphics[2][]{}%
  }%
  \providecommand\rotatebox[2]{#2}%
  \@ifundefined{ifGPcolor}{%
    \newif\ifGPcolor
    \GPcolortrue
  }{}%
  \@ifundefined{ifGPblacktext}{%
    \newif\ifGPblacktext
    \GPblacktextfalse
  }{}%
  \let\gplgaddtomacro\g@addto@macro
  \gdef\gplbacktext{}%
  \gdef\gplfronttext{}%
  \makeatother
  \ifGPblacktext
    \def\colorrgb#1{}%
    \def\colorgray#1{}%
  \else
    \ifGPcolor
      \def\colorrgb#1{\color[rgb]{#1}}%
      \def\colorgray#1{\color[gray]{#1}}%
      \expandafter\def\csname LTw\endcsname{\color{white}}%
      \expandafter\def\csname LTb\endcsname{\color{black}}%
      \expandafter\def\csname LTa\endcsname{\color{black}}%
      \expandafter\def\csname LT0\endcsname{\color[rgb]{1,0,0}}%
      \expandafter\def\csname LT1\endcsname{\color[rgb]{0,1,0}}%
      \expandafter\def\csname LT2\endcsname{\color[rgb]{0,0,1}}%
      \expandafter\def\csname LT3\endcsname{\color[rgb]{1,0,1}}%
      \expandafter\def\csname LT4\endcsname{\color[rgb]{0,1,1}}%
      \expandafter\def\csname LT5\endcsname{\color[rgb]{1,1,0}}%
      \expandafter\def\csname LT6\endcsname{\color[rgb]{0,0,0}}%
      \expandafter\def\csname LT7\endcsname{\color[rgb]{1,0.3,0}}%
      \expandafter\def\csname LT8\endcsname{\color[rgb]{0.5,0.5,0.5}}%
    \else
      \def\colorrgb#1{\color{black}}%
      \def\colorgray#1{\color[gray]{#1}}%
      \expandafter\def\csname LTw\endcsname{\color{white}}%
      \expandafter\def\csname LTb\endcsname{\color{black}}%
      \expandafter\def\csname LTa\endcsname{\color{black}}%
      \expandafter\def\csname LT0\endcsname{\color{black}}%
      \expandafter\def\csname LT1\endcsname{\color{black}}%
      \expandafter\def\csname LT2\endcsname{\color{black}}%
      \expandafter\def\csname LT3\endcsname{\color{black}}%
      \expandafter\def\csname LT4\endcsname{\color{black}}%
      \expandafter\def\csname LT5\endcsname{\color{black}}%
      \expandafter\def\csname LT6\endcsname{\color{black}}%
      \expandafter\def\csname LT7\endcsname{\color{black}}%
      \expandafter\def\csname LT8\endcsname{\color{black}}%
    \fi
  \fi
    \setlength{\unitlength}{0.0500bp}%
    \ifx\gptboxheight\undefined%
      \newlength{\gptboxheight}%
      \newlength{\gptboxwidth}%
      \newsavebox{\gptboxtext}%
    \fi%
    \setlength{\fboxrule}{0.5pt}%
    \setlength{\fboxsep}{1pt}%
\begin{picture}(4874.00,5102.00)%
    \gplgaddtomacro\gplbacktext{%
      \csname LTb\endcsname%
      \put(792,3255){\makebox(0,0)[r]{\strut{}$10^{-6}$}}%
      \put(792,3757){\makebox(0,0)[r]{\strut{}$10^{-4}$}}%
      \put(792,4259){\makebox(0,0)[r]{\strut{}$10^{-2}$}}%
      \put(792,4761){\makebox(0,0)[r]{\strut{}$10^{0}$}}%
      \put(1068,3035){\makebox(0,0){\strut{}$-6$}}%
      \put(1546,3035){\makebox(0,0){\strut{}$-4$}}%
      \put(2024,3035){\makebox(0,0){\strut{}$-2$}}%
      \put(2503,3035){\makebox(0,0){\strut{}$0$}}%
      \put(2981,3035){\makebox(0,0){\strut{}$2$}}%
      \put(3459,3035){\makebox(0,0){\strut{}$4$}}%
      \put(3938,3035){\makebox(0,0){\strut{}$6$}}%
      \colorrgb{0.02,0.81,0.47}%
      \put(1905,4259){\rotatebox{-80}{\makebox(0,0)[l]{\strut{}$\text{p} \rightarrow \text{a-p}$}}}%
      \colorrgb{0.82,0.21,0.80}%
      \put(2622,4259){\rotatebox{85}{\makebox(0,0)[r]{\strut{}a-p $\rightarrow$ p}}}%
      \csname LTb\endcsname%
      \put(49,4846){\makebox(0,0)[l]{\strut{}(a)}}%
      \colorrgb{0.00,0.00,0.00}%
      \put(2902,3375){\makebox(0,0)[l]{\strut{}$t_\text{PW}=\SI{100}{\nano\second}$}}%
    }%
    \gplgaddtomacro\gplfronttext{%
      \csname LTb\endcsname%
      \put(181,4046){\rotatebox{-270}{\makebox(0,0){\strut{}$P_\text{SW}$}}}%
      \put(2502,2705){\makebox(0,0){\strut{}$I_\text{Pulse}$ (a. u.)}}%
    }%
    \gplgaddtomacro\gplbacktext{%
      \csname LTb\endcsname%
      \put(792,704){\makebox(0,0)[r]{\strut{}$55$}}%
      \put(792,1100){\makebox(0,0)[r]{\strut{}$60$}}%
      \put(792,1496){\makebox(0,0)[r]{\strut{}$65$}}%
      \put(792,1891){\makebox(0,0)[r]{\strut{}$70$}}%
      \put(792,2287){\makebox(0,0)[r]{\strut{}$75$}}%
      \put(924,484){\makebox(0,0){\strut{}$0$}}%
      \put(1826,484){\makebox(0,0){\strut{}$0.02$}}%
      \put(2728,484){\makebox(0,0){\strut{}$0.04$}}%
      \put(3630,484){\makebox(0,0){\strut{}$0.06$}}%
      \colorrgb{0.87,0.09,0.12}%
      \put(1375,1733){\makebox(0,0)[l]{\strut{}parallel $\rightarrow$ anti-parallel}}%
      \colorrgb{0.00,0.38,0.68}%
      \put(1375,1258){\makebox(0,0)[l]{\strut{}anti-parallel $\rightarrow$ parallel}}%
      \csname LTb\endcsname%
      \put(49,2295){\makebox(0,0)[l]{\strut{}(b)}}%
    }%
    \gplgaddtomacro\gplfronttext{%
      \csname LTb\endcsname%
      \put(220,1495){\rotatebox{-270}{\makebox(0,0){\strut{}$\Delta$}}}%
      \put(2502,154){\makebox(0,0){\strut{}$\epsilon $ ($\%$)}}%
    }%
    \gplbacktext
    \put(0,0){\includegraphics{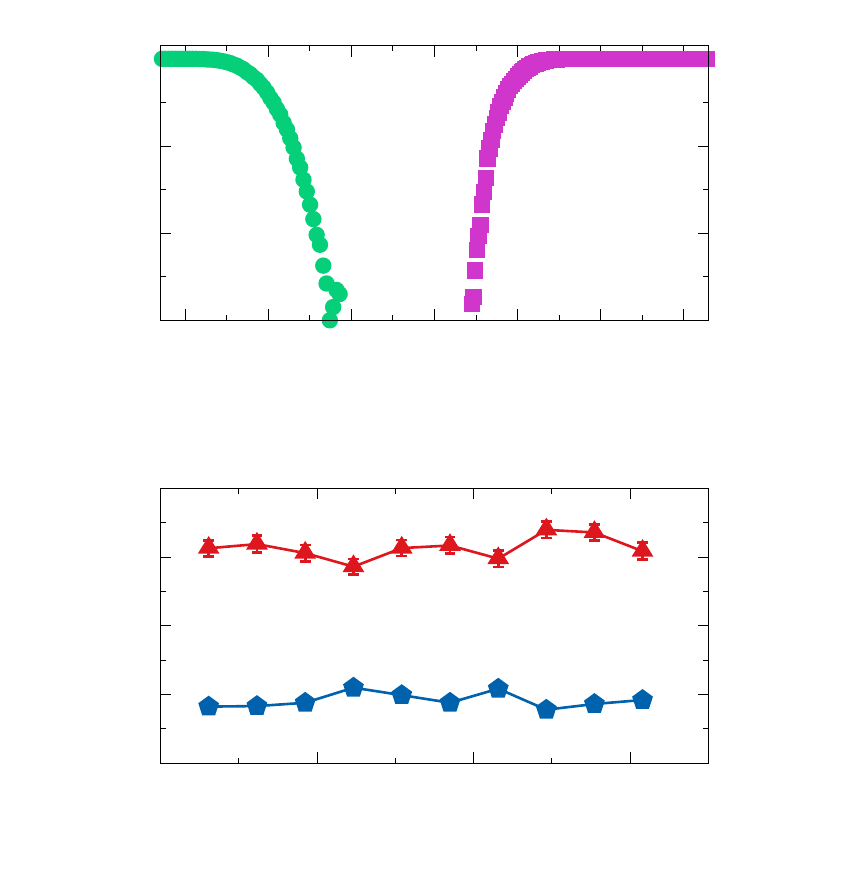}}%
    \gplfronttext
  \end{picture}%
\endgroup
\caption{\label{fig:Fig4} (a) For write error rate (WER) testing, the switching probability is measured as a function of pulse amplitude. The pulse length is $t_\text{PW}=\SI{100}{\nano\second}$. (b) Thermal stability factor $\Delta$, extracted from WER measurements, as a function of strain. The thermal stability does not show any variation as a function of strain within the measurement accuracy.}
\end{center}
\end{figure}
The thermal stability $\Delta$ is derived from write error rate (WER) measurements. The switching probability $P_{SW}$ is plotted as a function of current pulse amplitude $I_\text{Pulse}$ in Fig.~\ref{fig:Fig4}(a). We chose a pulse length of $t_\text{PW}=\SI{100}{\nano\second}$ and measured deep error rates down to $P_\text{SW}=10^{-6}$. The thermal stability $\Delta$ is calculated according to~\cite{Butler2012}:
\begin{align*}
\Delta = -\log(P_\text{SW}^{0})+\log(t_\text{PW}) \, ,
\end{align*}
where $P_\text{SW}^{0}$ is the extrapolated switching probability at $I_\text{Pulse}=0$.\\
Figure~\ref{fig:Fig4}(b) shows $\Delta$, averaged for 63 devices, as a function of strain. $\Delta$ is constant for all strain values tested. It should be noted that the noise in this measurement is on the order of $\SI{2}{\percent}$ of the mean value. Thus, if the change in $\Delta$ is of the same order of magnitude as the change in $\mu_0H_\text c$ or $I_{50\%}^{\SI{100}{\nano\second}}$, it will not be detectable by this method. \\
In conclusion, we have measured MTJ performance parameters under the application of mechanical strain. The strain was applied in a 4-point bending geometry, where the strain is constant over a large substrate area. It is found that the TMR ratio as well as the thermal stability in the devices under test do not change as a function of strain within the measurement accuracy. The coercive field and the switching current decrease by approximately $\SI{2}{\percent}$ over the whole range of applied strain. A thinner free layer in our devices with PMA might contribute to the quantitative difference in the strain dependence of the TMR seen in previous work~\cite{Loong2015,Loong2016}. The result reported here has significant implications for the manufacturability of STT-MRAM, as strain is often the result of device encapsulation or CMOS passivation processes that are determined by BEOL requirements.
\FloatBarrier
\vspace{1 cm}
We thank Robert Beach and Volodymyr Voznyuk for fruitful discussions.

\FloatBarrier

\end{document}